\documentstyle[12pt]{article}
\begin{document}
\input epsf
\noindent
Comment on ``Generalized Dynamic Scaling for Critical Relaxation"\\[2mm]
[Phys. Rev. Lett. 77, 679 (1996)]\\[4mm] 
In the work by Zheng \cite{zheng}
the temporal evolution of a spin system that is quenched from a
high-temperature initial state to the critical point $T_c$ is
studied by means of scaling analyses and
Monte Carlo simulations.
Special attention is paid to the time dependence and scaling behavior
of the order-parameter $m(t)$,
when the system -- the concrete case studied was the 3-state Potts
model -- is quenched from a state with a finite 
initial magnetization $m_0$.

The analysis of Ref.\,\cite{zheng} is built upon
previous work \cite{own,monte,liea} where the
influence of a {\it small} $m_0$
on relaxation processes in finite systems
was studied.
When, after the quench, the time-dependent
correlation length $\xi(t)\sim t^{1/z}$ (where $z$ denotes
the dynamic equilibrium exponent) has grown much larger
than macroscopic scales, e.g. the lattice spacing $a$, the order
parameter $m(t,m_0)$ should, due to the loss of intrinsic scales,
exhibit {\it universal} behavior. Especially
for $m_0\to 0$ in a system of finite linear size $L\,(\gg a)$
the scale invariance can be expressed in the form
\begin{equation}\label{scal}
m(t,\,L,\,m_0)=b^{-\beta/\nu}\,m(t\,b^{-z},L\,b^{-1},\,m_0\,b^{\,x_0})\>.
\end{equation}
The exponent $x_0$ in this relation, the scaling dimension
of the initial magnetization, is an independent critical
exponent that in general is different from the scaling dimension
$\beta/\nu$ of the equilibrium magnetization \cite{jans,own}.
By mapping Monte Carlo profiles
from systems of different size onto each other and
by employing Eq.\,(\ref{scal}), it was
possible to estimate the new exponent $x_0$ \cite{monte,liea}.

For larger $m_0$, Eq.\,(\ref{scal}) as it stands
becomes certainly wrong
since $m_0$ saturates. Nevertheless one would expect 
that in a stage
where $\xi(t)$ has grown much larger than $a$
the temporal evolution of $m(t)$ is still universal and should
exhibit scaling.
To verify this, Zheng
allows in the last argument on the right-hand side
of (\ref{scal}) for a general function $m_0'(m_0)$.
Similar to the
notation for small $m_0$, this function is written in the form $m_0'=\, 
b^{x_0(b,m_0)}\,m_0$, and $x_0(b,m_0)$
is determined as a function of $m_0$. The results presented
in Figs.\,1, 2, and 3 of Ref.\,\cite{zheng}
prove that scaling occurs for general $m_0$, and 
there is no doubt on the correctness of these results.

What we want to point out in this Comment is the following:
In order to take into account the initial state in the scaling
analysis the field that has to enter the description is actually
the {\it magnetic field} $H_0$ that is imposed in the initial state (and
switched off at $t=0$) and {\it not} the initial magnetization.
Only for {\it small} $H_0$ one can
effectively replace $H_0$ by $m_0$ for the initial state is paramagnetic
and responds linearly on small magnetic fields. As a consequence, it was
not necessary to distinguish between $H_0$ and $m_0$ in most of the
earlier work\cite{own,monte}. Now, the original $x_0$ in (\ref{scal})
is actually the
scaling dimension of $H_0$ -- we call it $y_0$ in the following --,
and the distinction becomes important
in a regime where the response is not linear anymore. 
As a result, Eq.\,1 should be replaced by
\begin{equation}\label{scal2}
m(t,\,L,\,H_0)=b^{-\beta/\nu}\,m(t\,b^{-z},L\,b^{-1},\,H_0\,b^{y_0})\>,
\end{equation}
and especially $y_0$ should be simply 
a {\it constant}
for all $H_0$.
Assuming the latter and using that
$m_0=\tanh(H_0)$ in our high-temperature initial state, the function
$x_0(b,m_0)$ occurring in (\ref{scal}) can even
be calculated. The result is given by
\begin{equation}\label{x0}
x_0(b,\,m_0)=\frac{1}{\log b}\left\{\log\left[\tanh\left(b^{y_0}\,
\mbox{artanh}\,m_0\right)\right]-\log m_0\right\}\>.
\end{equation}

To corroborate this by Monte Carlo simulations,
we calculated a number of
magnetization profiles for the $d=2$ Ising
model for lattices with sizes $40^2$ and $80^2$, such that $b=2$,
and with periodic boundaries.
Further we used the heat-bath algorithm and sequential updating.
In order to rescale the profiles
we adopted the literature value $z=2.17$\cite{grass} and used the exact
exponent $\beta/\nu=1/8$.
In Fig.\,1 we show 
$y_0$ (squares) and $x_0$ (full circles)
for $b=2$. The errors were estimated. The value 
$x_0=y_0\simeq 0.54$ for $m_0=0$ is derived 
from the short-time exponent $\theta=
(x_0-\beta/\nu)/z\simeq 0.19$ \cite{liea,grass}.
Further, we have obviously $x_0=0$ for $m_0=1$,
whereas $y_0$ can not be determined at this point since
$H_0=\infty$. For comparison we also show the result of
Eq.\,(\ref{x0}) for $x_0$ (solid line), which is consistent
with the numerical data.
Most importantly, our results for $y_0$ are
consistent with a constant (dashed line) for all $H_0$. 

\begin{figure}[t]
\def\epsfsize#1#2{0.53#1}
\hspace*{2.5cm}\epsfbox{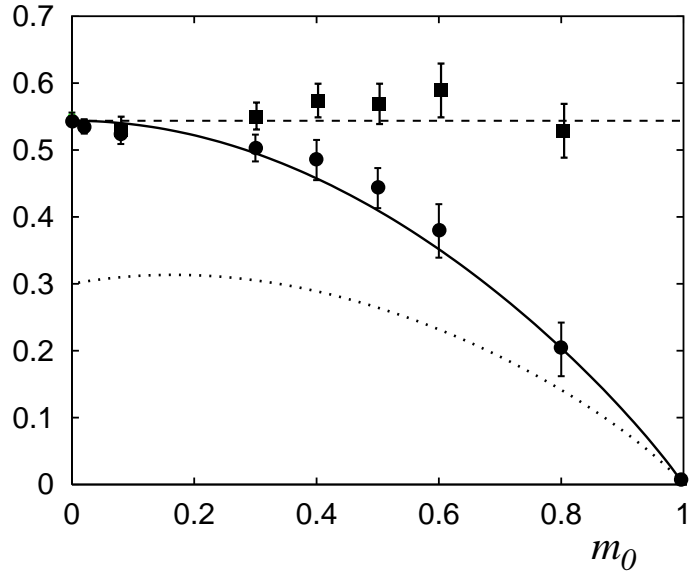}\\[0mm]
\caption{The scaling dimensions $x_0$ (full circles) and
$y_0$ (squares) as a function of $m_0$ from lattices of
size $40^2$ and $80^2$. The solid line represents the result
of Eq.\,(3) for $x_0(2,m_0)$, and the dotted line
is the corresponding curve for the 3-state Potts model.}
\end{figure}

Eq.\,(\ref{x0}) can straightforwardly be generalized for
the $p$-state Potts model. The result for $p=3$ with
$x_0(2,0)=0.3$ is also shown in Fig.\,1 (dotted line).
The curve is qualitatively consistent
with the data of Ref.\,\cite{zheng}, however, it does not provide
a quantitatively correct fit. This discrepancy could be related
to the fact that in Zheng's simulations the initial state
is not a thermal equilibrium state but has a sharp $m_0$.
We do not believe that for the general Potts model (\ref{scal2}) fails.
Our simulations always started from a thermal
equilibrium state.

Let us finally remark that a situation in many
respects analogous to the one discussed above is met in
{\it equilibrium} systems
with a planar surface (semi-infinite geometry)
at bulk criticality. To the high-temperature initial state
in the dynamical problem
corresponds a surface that strongly suppresses the order, and
the analogue to $H_0$ is a magnetic field $H_1$ that is restricted
to a microscopic region near the surface. As discussed in
Ref.\,\cite{nslro} it is of particular importance to keep
fields and densities distinguished in the context of
surface critical phenomena, and
much can be learned from there 
for nonequilibrium critical dynamics (and vice versa).

\hspace*{1cm}
U. Ritschel and P. Czerner\\
\hspace*{1cm}
Fachbereich Physik\\
\hspace*{1cm}
Universit\"at Essen\\
\hspace*{1cm}
45117 Essen (Germany)
\end{document}